\documentclass[twocolumn,showpacs,preprintnumbers,amsmath,amssymb,groupedaddress,superscriptaddress]{revtex4}

\usepackage{graphicx}
\usepackage{dcolumn}

\begin{document}


\title{Coherent population trapping resonances with linearly polarized light \\
for all-optical miniature atomic clocks}

\author {Sergei A.~Zibrov} \email  {szibrov@yandex.ru}
\affiliation{P.N.~Lebedev Physical Institute RAS, 117924 Moscow, Russia} \affiliation{Moscow State Engineering
Physics Institute, 115409 Moscow, Russia}
\author {Irina ~Novikova} \email  {inovikova@physics.wm.edu}
\affiliation{Department of Physics, College of William\&Mary, Williamsburg, Virginia 23185, USA}
\affiliation{Harvard-Smithsonian Center for Astrophysics, Cambridge, Massachusetts, 02138, USA}
\author {David F.~Phillips}
\affiliation{Harvard-Smithsonian Center for Astrophysics, Cambridge, Massachusetts, 02138, USA}
\author {Ronald L.~Walsworth}
\affiliation{Harvard-Smithsonian Center for Astrophysics, Cambridge, Massachusetts, 02138, USA}
\affiliation{Department of Physics, Harvard University, Cambridge, Massachusetts, 02138, USA}
\author {Alexander S.~Zibrov}
 \affiliation{Department of Physics, Harvard University, Cambridge, Massachusetts, 02138, USA}
\affiliation{P.N.~Lebedev Physical Institute RAS, 117924 Moscow, Russia} \affiliation{Moscow State Engineering
Physics Institute, 115409 Moscow, Russia}
\author {Vladimir L.~Velichansky}
\affiliation{P.N.~Lebedev Physical Institute RAS, 117924 Moscow, Russia} \affiliation{Moscow State Engineering
Physics Institute, 115409 Moscow, Russia}
\author {Alexey V.~Taichenachev}
\affiliation{Institute of Laser Physics SB RAS, Novosibirsk
630090, Russia} 
\affiliation{Novosibirsk State Technical University, Novosibirsk 630092, Russia}
\author {Valery I.~Yudin} \email{viyudin@mail.ru}
\affiliation{Institute of Laser Physics SB RAS, Novosibirsk 630090, Russia} \affiliation{Novosibirsk State
Technical University, Novosibirsk 630092, Russia}\affiliation{Novosibirsk State University, Novosibirsk 630090,
Russia}

\date{\today}

\begin{abstract}
We present a joint theoretical and experimental characterization of the coherent population trapping (CPT)
resonance excited on the D$_1$ line of $^{87}$Rb atoms by bichromatic linearly polarized laser light. We observe
high-contrast transmission resonances (up to $\approx 25\%$), which makes this excitation scheme promising for
miniature all-optical atomic clock applications. We also demonstrate cancellation of the first-order light shift
by proper choice of the frequencies and relative intensities of the two laser field components. Our theoretical
predictions are in good agreement with the experimental results.
\end{abstract}


\maketitle

\section{Introduction}

There is great interest in the development of miniature (chip-scale) atomic clocks, magnetometers and other
metrology tools. One of the promising approaches utilizes coherent population trapping (CPT) resonances: by
coherent interaction with two resonant optical fields in a $\Lambda$ configuration, atoms can be prepared in a
non-interacting quantum superposition of two hyperfine ground states (a ``dark'' state), which results in
enhanced transmission~\cite{vanier05apb}. The dark state is very sensitive to detuning of the frequency
difference of the two optical fields from the hyperfine splitting of the atomic states, which allows using the
CPT effect for precision sensing. In recent years atomic frequency standards based on CPT resonances with
fractional frequency stability of 10$^{-12}$ or better have been demonstrated
\cite{KitchingIEEE00,Finland,vanier05apb}. Further improvements in stability require realization of CPT
resonances with optimized parameters, such as larger amplitude and contrast and smaller linewidth. In addition,
any sensitivities to environmental changes must be minimized. In particular, light shifts (i.e., spectra shifts
in the transmission peak with total light power) can limit the frequency stability of CPT-based devices.

\begin{figure}[h]
  \includegraphics[width=0.9\columnwidth]{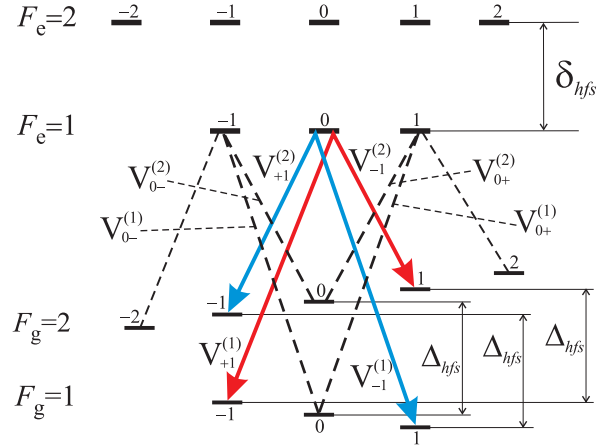}
  \caption{\emph{(Color online)} Light-induced two-photon transitions in a bichromatic field from $F_g=1,2$ to $F_e=1$
  for different variants of optical polarization.
} \label{fig1}
\end{figure}
\suppressfloats

Miniature atomic clocks typically use a single diode laser (usually a vertical cavity surface-emitting laser, or
VCSEL) phase-modulated at an appropriate frequency to obtain two light fields, required for CPT resonances. As a
result, practical operational schemes are restricted to light fields of the same polarization. In the case of
circularly polarized light fields, for example, this restriction leads to significant reduction of CPT resonance
amplitude, since large fraction of the atomic population becomes trapped in the ``end'' Zeeman states with
maximum value of the angular momentum ($m_F=F$ or $m_F=-F$), and not in the desired magnetic field - insensitive
states with $m_F=0$~\cite{jauPRL04}. To mitigate this problem, a variety of CPT
schemes~\cite{TaichenachevJETP04,JauPRL04,ZanonPRL05,shahOL07} as well as related
coherent resonances~\cite{vukicevicIEEE00,zibrov05pra,Nlightshifts} have been proposed and studied by different groups. 

In this paper we present a CPT interaction scheme that takes advantage of the unique level combination of alkali
metal atoms with nuclear spin $I$=3/2 to realize high-contrast magneto-insensitive CPT resonances with two
optical fields of the same linear polarization ({\em lin}$||${\em lin} configuration). This scheme, initially
proposed in Ref.~\cite{linlin05}, is possible when two ground states with total angular momentum $F_g$=1,2 are
coupled only through an excited state with total angular momentum $F_e$=1. In a vapor cell this situation is
realized only for the D$_1$ line of $^{87}$Rb when the exited-state hyperfine levels are spectrally resolved.
Here, we present an experimental and theoretical study of this {\em lin}$||${\em lin} CPT scheme, and
demonstrate the cancellation of the first-order light shift by proper choice of the frequencies and relative
intensities of the two optical fields, which can reduce the sensitivity of resonance parameters to laser
intensity fluctuations.

\section{Theoretical analysis}

Observation of CPT resonances requires two optical fields ${\bf E}_{1,2} e^{[-i(\omega_{1,2}t-k_{1,2} z)]}$
tuned to two-photon resonance with the hyperfine transition of an appropriate atom:
$\omega_{1}-\omega_{2}=\Delta_{hfs}$. In ${}^{87}$Rb, for example, the hyperfine splitting is $\Delta_{hfs}
\simeq 6.835$~GHz. For miniature atomic clock applications, both fields are produced by a single phase-modulated
laser (typically a VCSEL) and have the same polarization. Traditional CPT clocks use circularly polarized laser
output~\cite{vanier05apb}. In this case the magnetic field-insensitive CPT resonance is formed between $m_F=0$
Zeeman sublevels [(0)-(0) clock transition]. For example, for $\sigma_+$-polarized laser fields $E^{(1,2)}_{+}$,
the corresponding CPT dark state is:
\begin{eqnarray}\label{D_circ}
&&|dark_{circ}\rangle= \mathcal{N}\times\\
&&\left\{|F_g=1,m=0\rangle -\frac{V^{(1)}_{0+}E^{(1)}_{+}}{V^{(2)}_{0+}E^{(2)}_{+} }\,|F_g=2,m=0\rangle\right\},
\nonumber
\end{eqnarray}
where $V^{(1,2)}_{0}$ are the transition matrix elements  for $|F_g=1,2,m=0\rangle \rightarrow |F_e,m=1\rangle$,
and $\mathcal{N}$ is a normalization coefficient. The disadvantage of such a scheme is the existence of the
noninteracting trap state $|trap_{circ}\rangle = |F_g=2,m=2\rangle$: since it is decoupled from both circularly
polarized optical fields, a large portion of the atomic population is optically pumped in the trap state. This
effectively reduces the density of interacting atoms, resulting in lower CPT resonance contrast.

Application of linearly (or in general elliptically) polarized optical fields can eliminate such trap states.
However, such polarizations simultaneously deteriorate the ground-state coherence formed between  two the
magnetic field-insensitive sublevels $m=0$ in the presence of a longitudinal magnetic field. In this case, two
opposite circularly polarized components of the optical fields simultaneously form two $\Lambda$-systems between
the $m=0$ ground states, through both $|F_e,m=\pm1\rangle$ excited states. The matrix elements for
$|F_g=2,m=0\rangle \rightarrow |F_e,m=\pm1\rangle$ are equal and opposite $V^{(2)}_{0+} = -V^{(2)}_{0-}$, but
those for the transitions $|F_g=1,m=0\rangle \rightarrow |F_e,m=\pm1\rangle$ are the same $V^{(1)}_{0+} =
V^{(1)}_{0-}$. As a result, the dark states for each $\Lambda$ system are different, as can be seen from
Eq.~\ref{D_circ}. In particular, for linearly polarized light, when the intensities of the two circular
components are the same, the ground-state coherence on the (0)-(0) clock transition disappears completely, since
the dark state for one $\Lambda$ system becomes the bright (interacting) state for the other.

${}^{87}$Rb atoms, with the nuclear spin of {\em I}=3/2, offer a unique possibility to form high contrast
magneto-insensitive CPT dark resonances using a different combination of Zeeman sublevels with arbitrary
polarized light fields even for the condition of reduced or zero coherence on the (0)-(0) clock transition.
Possible $\mathrm{D}_1$ optical transitions from $F_g=1,2$ ground state to the $F_e=1$ excited state resulting
from the interaction of two elliptically polarized optical fields are shown in Fig.~\ref{fig1}. Solid arrows
show two possible two-photon $\Lambda$ resonances that create atomic coherence between $|F_g$=1,{\em
m}=$-1\rangle$, $|F_g$=2,{\em m}=$+1\rangle$ and between $|F_g$=1,{\em m}=$+1\rangle$, $|F_g$=2,{\em
m}=$-1\rangle$ [further in the text we refer to these as ($-$1)--($+$1) and ($+$1)--($-$1) resonances]. Both of
these $\Lambda$-schemes are excited through a common excited state $|F_e$=1,{\em m}=$0\rangle$. Most
importantly, the gyromagnetic $g_F$-factors of the two ground-state hyperfine levels in $I=3/2$ alkali metal
atoms have equal absolute value and opposite signs (neglecting the small contribution of nuclear spin). Thus
both ground states in each of these $\Lambda$ systems shift by an equal amount when a weak
longitudinal magnetic field is applied. 
%
As a result a CPT resonance occurs at the unshifted ``clock'' frequency $\omega_1-\omega_2 = \Delta_{hfs}$ via
two dark states $|dark_\pm\rangle$:
\begin{eqnarray}\label{D_pm}
&&|dark_\pm\rangle=\mathcal{N}_{\pm}\times\\
&&\left\{|F_g=1,m=\mp 1\rangle -\frac{V^{(1)}_{\pm 1}E^{(1)}_{\pm 1}}{V^{(2)}_{\mp 1}E^{(2)}_{\mp 1}
}\,|F_g=2,m=\pm 1\rangle\right\} .\nonumber
\end{eqnarray}
Here $V^{(1,2)}_{\pm 1}$ are the transition matrix elements shown in Fig.1, and $\mathcal{N}_{\pm}$ are
normalization constants. In this configuration there are no ``trap'' states when the Zeeman levels are not
degenerate. Hence under CPT conditions the atomic population accumulats only in the coherent dark states,
resulting in higher CPT resonance contrast.

While the states (\ref{D_pm}) exist for arbitrary elliptical polarizations of the optical fields, a linearly
polarized laser output is preferable since the influence of nuclear spin leads to a small difference in the
$g$-factors for different hyperfine components of the ground state. As a result of this $g$-factor difference
the frequencies of the ($-1$)--($+1$) and ($+1$)--($-1$) two-photon $\Lambda$-resonances are also different in
finite magnetic field (for $^{87}$Rb this difference is 2.8 kHz/G), but their position is symmetric relative to
the hyperfine splitting frequency $\Delta_{hfs}$. Thus, for the {\em lin}$||${\em lin} excitation scheme, this
small difference in the two-photon resonance frequencies leads only to the broadening of the resonance, but not
to a frequency shift (in linear Zeeman effect approximation). At stronger magnetic field the CPT resonances from
the two $\Lambda$ systems become resolvable; the dip between the two CPT peak may be used as a
``pseudo-resonance" for atomic clocks as well, although its characteristics look less
promising~\cite{matisovPRA05,zibrovJETPLett05}. 

It is important to note that if the ground states $|F_g$=1,{\em m}=$\pm 1\rangle$ and $|F_g$=2,{\em m}=$\pm
1\rangle$ are coupled to the excited states $|F_e$=2,{\em m}=$\pm 2\rangle$, then the dark state given by
Eq.(\ref{D_pm}) are degraded~\cite{watabeAO09}. Thus, the key requirement to the observation of high-contrast
{\em lin}$||${\em lin} CPT resonances is good spectral resolution of the excited state $F_e$=1. The D1 line of
$^{87}$Rb is the most promising candidate for this scheme since its excited state hyperfine splitting (812~MHz)
exceeds the Doppler broadening at room temperature, allowing its use in vapor cells and hence in miniature
atomic clocks. For all other $I=3/2$ alkali atoms ($^{7}$Li, $^{23}$Na, $^{39,41}$K), and also for the D${}_2$
line of $^{87}$Rb, good spectral resolution of the excited state is possible only in the case of laser cooled
atoms or in a collimated atomic beam.
Even for the $^{87}$Rb D1 line the realization of a miniature {\em lin}$||${\em lin} CPT atomic clocks will
require careful optimization of the buffer gas pressure to increase the diffusion time of atoms in the laser
beam and, hence, the dark state lifetime without severe degradation of the resonance amplitude due to the
reduced spectral resolution of the pressure-broadened excited states. In particular, there is a potential
difficulty in using this scheme for chip-scale atomic clocks that typically use very high buffer gas pressures,
which broaden the optical resonances beyond the usual Doppler linewidth. This difficulty can be circumvented,
however, through the addition of an anti-relaxation wall coating to a vapor cells with lower buffer gas
pressure~\cite{budkerPRA06coating}.

\section{\label{sec:level1}Experimental procedure}

\begin{figure}[t]
  \includegraphics[width=0.45\textwidth]{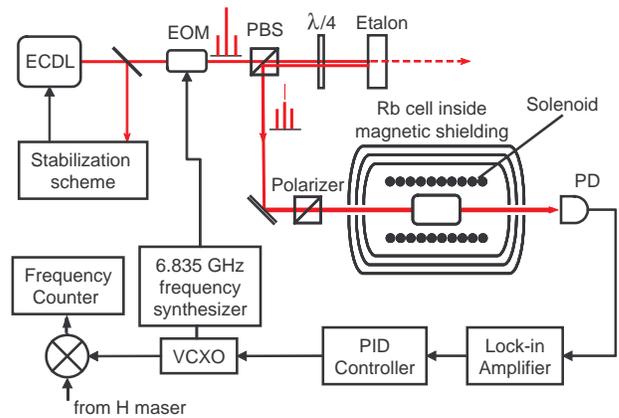}
  \caption{\emph{(Color online)} Schematic of the experimental setup. See text for abbreviations.}\label{fig2}
\end{figure}

A schematic of the experimental setup is shown in Figure~\ref{fig2}. To produce the required resonant
bichromatic field the output of an external cavity diode laser (ECDL) was externally phase-modulated at the
${}^{87}$Rb hyperfine frequency $6.835$~GHz by an electro-optical modulator (EOM).  The laser frequency was
stabilized to a saturation spectroscopy resonance in a vacuum reference cell, with the drive field [unmodulated
carrier] tuned in the vicinity of the $5S_{1/2} F_g=2 \leftrightarrow 5P_{1/2} F_e=1$ transition, and the probe
field [first high-frequency modulation sideband] resonant with the $5S_{1/2} F_g=1 \leftrightarrow 5P_{1/2}
F_e=1$ transition.
The two-photon detuning between the probe and drive fields was precisely controlled by varying the output
frequency of the microwave synthesizer providing the EOM modulation signal.
The intensity ratio $I_1/I_2$ between probe and drive fields [sideband/carrier ratio] was adjusted up to $40~\%$
by changing the power of the microwave signal sent to the EOM. To achieve higher intensity ratios we used a
thermo-stabilized quartz Fabry-Perot etalon with free spectral range of 20~GHz, finesse of 30, and maximum
transmission of 70$\%$, temperature-tuned to partially transmit the strong drive field. The beam reflected from
the etalon contained the unaffected first order sidebands and an attenuated drive field. The attenuation was
controlled by fine-tuning the transmission resonance of the etalon via temperature. The reflected beam was
collimated to a diameter of $7$~mm and linearly polarized by a quartz polarizer before entering the Rb vapor
cell.

All experimental measurements were recorded using a cylindrical
Pyrex cell (diameter~=~2.5~cm, length~=~4~cm) containing
isotopically enriched $^{87}$Rb and 4~Torr of Ne buffer gas to
increase the Rb interaction time with the laser beam. The cell was
mounted inside three-layers of magnetic shielding and maintained
at a constant temperature 40$^\circ$~C. To separate the
magneto-insensitive ``clock'' resonances ($|F_g=1,
m=-1\rangle\leftrightarrow |F_g=2, m=+1\rangle$ and $|F_g=1, m=+1
\rangle\leftrightarrow |F_g=2, m=-1\rangle$) from the
field-dependent resonances, a uniform longitudinal magnetic field
of approximately $180$~mG was applied using a solenoid mounted
inside the innermost magnetic shield. A photodetector (PD) placed
after the cell was used to measure the transmitted optical signal.
%
%
CPT resonance parameters such as linewidth, amplitude, etc., were determined for each set of experimental
conditions.

To continually monitor the CPT resonance frequency the EOM modulation frequency was locked to the CPT resonance
using phase-sensitive detection. A slow frequency modulation at $f_m = 400$~Hz was superimposed on the
$6.835$~GHz microwave synthesizer output; then the photodetector signal was demodulated at $f_m$ using a lock-in
amplifier. The resulting error signal was fed back to lock the frequency of the voltage controlled crystal
oscillator (VCXO) that controlledthe synthesizer. The frequency of the locked VCXO was measured by beating it
with a stable reference derived from a hydrogen maser.

\section{\label{sec:level1}Experimental results}

 \begin{figure}[t]
 \begin{center}
 \includegraphics[width=1.0\columnwidth]{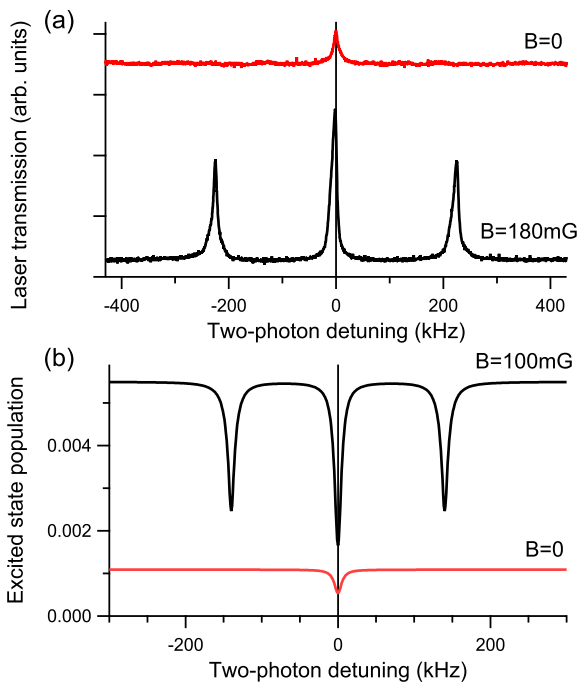}
  \caption{\label{Fig:SampleRes}
\emph{(Color online)} (\emph{a}) Examples of measured CPT resonances at $B=0$ and $B=180$~mG. Laser intensity
before cell is $8.4$~mW/cm$^2$.
 (\emph{b})
 Calculated excited-state population $\pi_e$
 versus two-photon detuning at $B=0$ and $B=100$~mG. Total resonant intensity $I_1+I_2 = 10$~mW/cm$^2$. }
 \end{center}
 \end{figure}

Fig.~\ref{Fig:SampleRes}(a) shows typical measured CPT resonances in the {\em lin}$||${\em lin} configuration.
The strong peak at zero two-photon detuning is from the magneto-insensitive ($-$1)--($+$1) and ($+$1)--($-$1)
CPT resonances; the two smaller side peaks correspond do the magnetic field-sensitive resonances formed between
the $|F_g=2, m=\pm 2 \rangle$ and $|F_g=1, m=\pm 0 \rangle$ states. Note that when the Zeeman levels are
degenerate (at $B=0$), coupling to the trap states $|F_g=2, m=\pm 2 \rangle$ significantly reduces the amplitude
and contrast of the CPT resonance (top line). A small magnetic field eliminates coupling to these trap states,
resulting in much higher resonance contrast.
Fig.\ref{Fig:SampleRes}(b) shows the calculated dependence of the total excited state populations $\pi_e$ as
function of two-photon detuning. These calculated spectra can be directly compared to the measured CPT
resonances, since $\pi_e$ is proportional to the laser field absorption~\cite{DR_theory}; small discrepancies
arise from residual magnetic field gradients.
\begin{figure*}[t]
  \includegraphics[width=1.5\columnwidth]{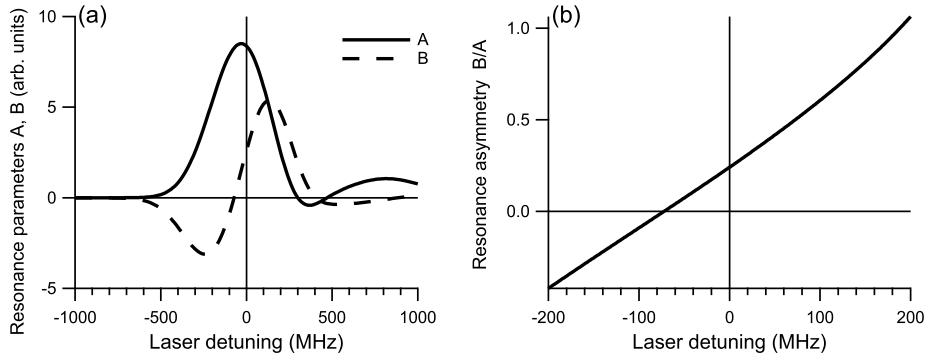}
  \caption{\textit{(a)} Calculated values of the lineshape parameters $A$ and $B$ [Eq.~(\ref{LLc})] that describe
  the symmetric and antisymmetric components of a CPT resonance as a function of laser detuning $\delta_L$ from the $F_e=1$ level.
  In these calculations: sideband/carrier ratio $I_1/I_2 = 0.5$; homogeneous collisionally-broadened optical linewidth
  $\gamma=20$~MHz; ground-state relaxation rate $\Gamma=500$~Hz, and the laser beam diameter $6.9$~cm.
  \textit{(b)} Calculated CPT resonance asymmetry $B/A$ as a function of laser detuning $\delta_L$ for $I_1/I_2 =
  0.5$, $\gamma=20$~MHz, $\Gamma=500$~Hz, beam diameter $6.9$~cm.
  CPT resonance becomes symmetric ($B=0$) at $\delta_L=-72$~MHz.
  }\label{Fig:res_assim}
\end{figure*}

We modeled each CPT-resonance lineshape as a function of the
two-photon (Raman) detuning $\delta_R$ by a generalized Lorentzian
\cite{KnappeAPB03,DR_theory}:
\begin{equation} \label{LLc}
I(\delta_R) =I_{bg} + \frac{A\,(\widetilde{\gamma}/2)^2} {(\widetilde{\gamma}/2)^2+(\delta_R-\delta_0)^2} +
\frac{B\,(\delta_R-\delta_0)\,\widetilde{\gamma}/2} {(\widetilde{\gamma}/2)^2+(\delta_R-\delta_0)^2},
\end{equation}
where $\widetilde{\gamma}$ is the width of the CPT resonance, $\delta_0$ is the resonance shift, $A$ and $B$
characterize the amplitudes of the symmetric and anti-symmetric Lorentzian components,  and $I_{bg}$ is the
total transmission away from CPT resonance (background transmission). See the Appendix for details. $A,B$,
$\widetilde{\gamma}$, and $\delta_0$ are functions of the optical field amplitudes and one-photon detuning
$\delta_L$; their values were extracted by fitting experimentally recorded or numerically calculated CPT
resonances to Eq.~\ref{LLc}. Fig.~\ref{Fig:res_assim}(a) shows the calculated variation of $A$ and $B$ when the
laser frequency is swept across the optical transition $F_g=2 \rightarrow F_e=1$ for typical experimental
conditions: one sees that the symmetric Loretzian component $A$ reaches its maximum near the optical resonance;
whereas the antisymmetric component $B$ is largest for $\delta_L \approx \pm 200$~MHz. This resonance asymmetry
can lead to shifts in measured resonance position if a phase-sensitive detection method is
used~\cite{phillipsJOSAB04}. Thus, it is desirable to operate a CPT atomic clock under conditions that lead to a
symmetric CPT resonance. Fig.~\ref{Fig:res_assim}(b) shows that for the {\em lin}$||${\em lin} configuration,
CPT resonances become symmetric ($B=0$) when the laser frequency is appropriately red-detuned from the $F_e=1$
excited state. We found a similar value for the optimal detuning experimentally. Note, that this detuning also
corresponds to the maximum of the the amplitude for the symmetric component $A$, making this operational
conditions attractive for clock operation. 

 \begin{figure*}[t]
 \begin{center}
     \includegraphics [width=1.5\columnwidth] {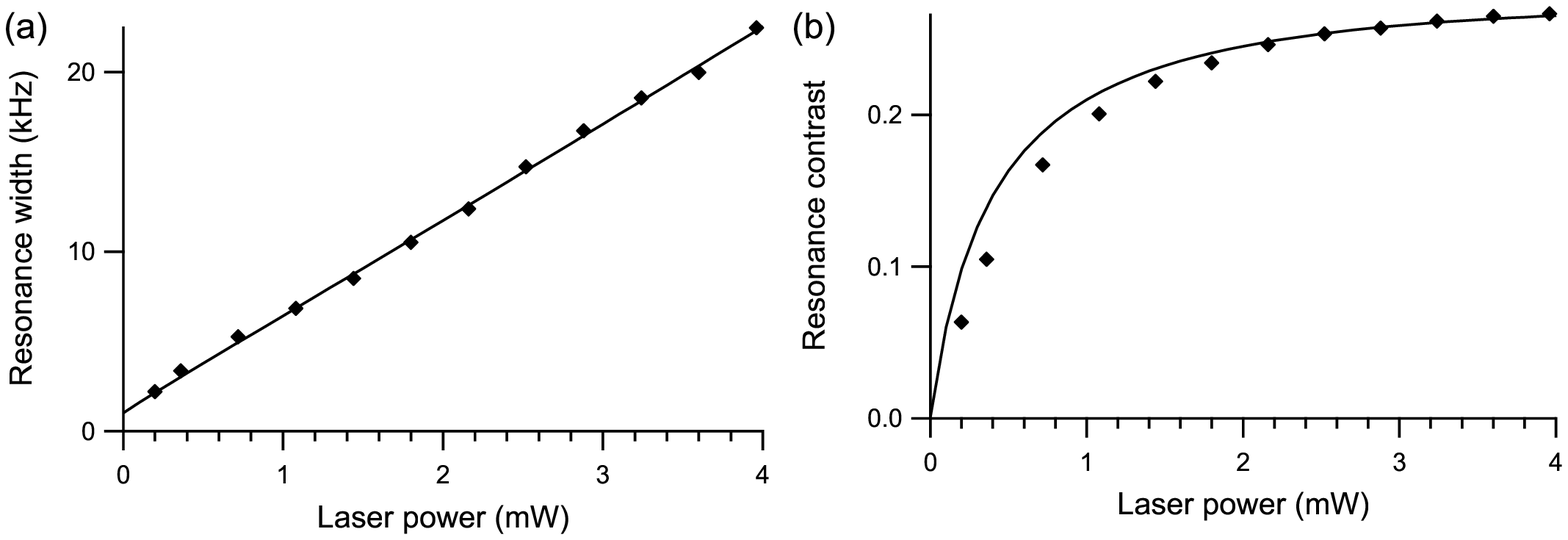}
 \caption{\label{Fig:CPTparameters}\emph{(a)}\emph{lin}$||$\emph{lin} CPT resonance linewidth $\,\widetilde{\gamma}$ and
 \emph{(b)} the resonance contrast $C$ (amplitude/background), for optimal laser detuning to achieve a symmetric resonance.
Experimental data points are shown as diamonds; solid lines represent numerical calculations.}
 \end{center}
 \end{figure*}
 %

Fig.~\ref{Fig:CPTparameters}(a) shows the measured dependance of the CPT resonance width as a function of total
laser power, for the optimal laser detuning ($\delta_L=-72$~MHz) to achieve a symmetric resonance. The results
of numerical calculations (solid lines) follow the experimental data quite closely; both show the expected
linear power broadening.
Fig.~\ref{Fig:CPTparameters}(b) shows experimental data and numerical calculations for the resonance contrast,
defined as the ration between the resonance amplitude and the off-resonant background transmission $C=A/I_{bg}$.
%
The contrast increases rapidly with the laser power, and then saturates at $C > 25$~\% for total incident laser
power $\geq 1$~mW. This maximum contrast greatly exceeds that achieved to date with traditional CPT clock
resonances, $C<4$\%~\cite{vanier05apb}, albeit with relatively large laser power and resonance width $\sim
10$~kHz.

The fractional frequency stability of a microwave oscillator locked to the CPT resonance is proportional to the
quality figure $q=C/\widetilde{\gamma}$ -- the ratio between the resonance contrast and width. Notably, $q$
characterizes the achievable stability for both shot noise-limited and intensity-fluctuation-limited atomic
clocks~\cite{Vanier03}. If the frequency stability of an atomic clock is limited by photon shot noise, the
expected Allan deviation, $\sigma(\tau)$ is given by~\cite{Vanier03a}:
\begin{equation}
  \sigma(\tau)=\frac{1}{4}\sqrt{\frac{\eta
  e}{I_{bg}}}\frac{1}{q\nu_0} \tau^{-1/2}
\label{e.stability}
\end{equation}
where $\nu_0=\Delta_{hfs}$ is the atomic reference frequency, $e$ is the electron charge, $\eta$ is the
photodetector sensitivity (measured optical energy per photoelectron), and $\tau$ is the integration time. For
the {\em lin}$||${\em lin} CPT resonances in our experiment (optimal laser detuning), the maximum $q\approx 3.2
\cdot 10^{-5}$/Hz value was measured at $\approx 800~\mu$W of laser power, which corresponds to a resonance
contrast $C\simeq 17\%$ and width $\widetilde{\gamma}\simeq 5$~kHz. This value implies a shot-noise-limited
frequency stability of $\sigma(\tau)\sim 2\cdot 10^{-14}{\tau}^{1/2}$. In practice, however, the frequency
stability will also be limited by residual intensity noise that depends strongly on the linewidth of the
laser~\cite{matisovIEEE09,matisovPRA09}. Recently, a table-top prototype atomic clock based on {\em lin}$||${\em
lin} CPT resonances with a short-term fractional frequency stability $2 \times 10^{-11} \tau^{-1/2}$
 has been demonstrated~\cite{novikovaJOSAB09}.


\section{\label{sec:shiftcompensation} Cancellation of first-order light shift}

\begin{figure}[t]
  \includegraphics[width=0.50\textwidth]{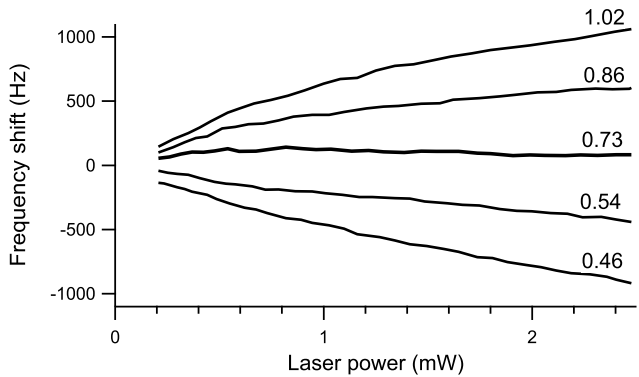}
  \caption{Measured CPT resonance frequency shift as a function of laser power
  for different sideband/carrier ratios (shown for each trace). Experimental conditions were otherwise the same
  as for Fig.~\ref{Fig:CPTparameters}. Ratio of $0.73$ corresponds to the light shift
  compensation condition (i.e., CPT resonance frequency minimally affected by laser power variation). }\label{shift_conc}
\end{figure}

\begin{figure}[t]
  \includegraphics[width=0.45\textwidth]{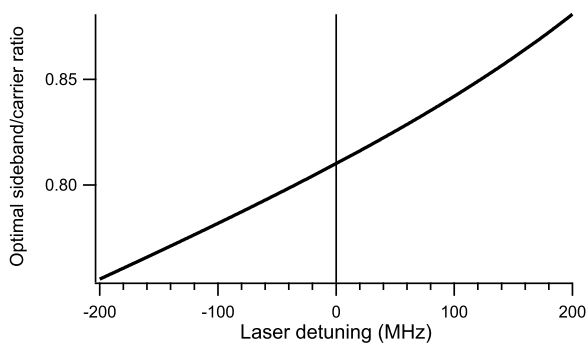}
  \caption{Calculated optimal sideband/carrier ratio,
for which light shift of CPT resonance position is minimally dependent on laser power, as a function of laser
detuning.}\label{rvsdelta_th}
\end{figure}

The light shift of a CPT resonance, i.e., changes in the maximum transmission  caused by changing laser
intensity~\cite{Mathur}, is another important characterization parameter for clock applications. Light shifts
deteriorate the stability of clocks, since they directly transfer any instabilities in the laser intensity into
variation of the measured clock frequency.
In the {\em lin}$||${\em lin} configuration the sensitivity of the CPT resonance position to laser intensity may
be
strongly reduced by choosing a proper intensity ratio between two optical fields. 
For a resonant $\Lambda$ system the main contribution to the CPT resonance light shift comes from the coupling
to non-resonant hyperfine levels, and in particular to the upper hyperfine level $F_e$=2, since
$\delta_{hfs}=812$~MHz$ \ll \Delta_{hfs}=6.835$~GHz. To motivate
first-order light shift cancellation we consider the two limiting
cases when one of the optical fields is much stronger than the
other. If $|E_1$$/$$E_2|$$\gg$1 the dark resonance position is
mostly affected by the light shift of the Zeeman sublevel
$|F_g$=1,{\em m}=$\pm 1\rangle$ caused by the strong $E_1$ field.
This Zeeman shift  is negative and proportional to $
-|V^{(1)}E_1|^2$$/\delta_{hfs}$, and results in a {\em positive}
shift of the CPT resonance. In the opposite case
$|E_1$$/$$E_2|$$\ll 1$, the main contribution is due to shift of
the Zeeman sublevel $|F_g$=2,{\em m}=$\pm 1\rangle$ $\propto
-|V^{(2)}E_2|^2$$/\delta_{hfs}$, which in this case leads to a
{\em negative} shift of the CPT resonance. Since in general the
frequency of the CPT peak will be determined by light shifts
created by both fields, it is reasonable to expect that by
changing the sideband/carrier ratio we can find a point of light
shift cancellation. Our calculations show that the compensation
condition depends only on the proper choice of the ratio of the
two fields and not on the absolute value of either optical
component. For the optimal one-photon laser detuning $\delta_L =
-72$~MHz, the calculations predict $I_1/I_2$=0.79 for light shift
compensation, which is in a good agreement with the experimentally
determined $I_1/I_2=0.73$ (see Fig.\ref{shift_conc}). The
compensation ratio depends slightly on laser detuning, as is shown
in Fig.\ref{rvsdelta_th}.

%
%

\section{Conclusion}

In this paper we presented theoretical and experimental characterization of magnetic field-insensitive CPT
resonances induced in $^{87}$Rb vapor by linearly polarized light. This excitation scheme allowed us to greatly
increase the contrast of magneto-insensitive CPT resonances. The dependance of the resonance amplitude,
linewidth, contrast, quality figure and frequency on total laser intensity was investigated. We also
demonstrated light-shift compensation by proper choice of the frequencies and intensity ratio of the resonant
optical fields. Our results suggest that {\em lin}$||${\em lin} CPT resonances may be good candidates for
miniature atomic clocks.

\section{Acknowledgements}
The authors are grateful to Eugeniy E. Mikhailov for useful discussions.  This work was supported by Smithsonian
Institution. I.N. acknowledges support from Jeffress Research grant J-847. A.V.T and V.I.Yu. were supported by
RFBR (grants 08-02-01108, 08-07-00127, 09-02-01151, 10-02-00406, 10-08-00844), Russian Academy of Science,
Presidium SB RAS, and by federal programs ``Development of scientific potential of higher school 2009-2010'' and
``Scientific and pedagogic personnel of innovative Russia 2009-2013''.

\section{\label{sec:DM}Appendix: density matrix calculations}

Our numerical calculations employ a model based on the standard density matrix approach (see e.g.,
\cite{DR_theory}). Under assumptions of low saturation and total collisional depolarization of the excited
state, the ground-state density matrix $\widehat{\sigma}_{gg}$ obeys the following equations:
\begin{eqnarray} \label{maineq}
&& \frac{d}{dt}\widehat{\sigma}_{gg} = -i\left[ \widehat{H}_{\rm eff}\widehat{\sigma}_{gg} -
\widehat{\sigma}_{gg}{\widehat{H}_{\rm eff}}^{\dagger} \right] +
\left(\frac{\pi_e}{\tau_e}+\Gamma\right)\,\frac{\widehat{\Pi}_g}{n_g}, \nonumber
\\
\label{n_cond} && {\rm Tr}\{\widehat{\sigma}_{gg}\} = 1,
\end{eqnarray}
where $\widehat{\Pi}_g $ is the unity matrix in the ground-state manifold, $n_g=2(2I+1)$ is the total number of
sub-states in the ground state, and $\pi_e$ is the total population of the excited state. The first source term
in Eq.(\ref{maineq}), inversely proportional to the excited-state radiative lifetime $\tau_e$, describes
isotropic repopulation of the ground-state sublevels due to spontaneous decay of the excited states. The second
source term, proportional to the ground-state relaxation rate $\Gamma$, describes the entrance of unpolarized
atoms into the laser beam due to diffusion and collisions.
 Hereafter we use the following shorthand notations:
$|e\rangle = |F_e,\,m_e\rangle$ with $m_e = -F_e,\ldots,F_e$, and $|i,m\rangle = |F_i,\,m\rangle$ with $m =
-F_i,\ldots,F_i$ ($i=1,2$). The non-Hermitian ground-state effective Hamiltonian has the form:
\begin{eqnarray} \label{Heff}
\widehat{H}_{\rm eff} &=& -\frac{\delta_{R}}2 \sum_{m}(|1,m\rangle \langle 1,m| - |2,m\rangle \langle 2,m|)
\\ &+&\widehat{H}_{B}
 + \langle\widehat{R}\rangle_v -i\frac{\Gamma}{2}\,\widehat{\Pi}_g. \nonumber
\end{eqnarray}
Here $\delta_{R}=\omega_1-\omega_2-\Delta_{hfs}$ is the two-photon (Raman) detuning; $\widehat{H}_{B}$ is the
magnetic Hamiltonian, describing the linear Zeeman splitting; the angle brackets $\langle\ldots\rangle_v$ denote
averaging over atomic velocities with a Maxwell distribution function. The two-photon excitation matrix for an
atom moving with a velocity $v$ can be written as
\begin{widetext}
\begin{equation} \label{Rmatr}
\widehat{R} = \sum_{i,j,e,m,m'}|i,m\rangle \frac{\langle i,m|{(\widehat{\bf d}\cdot {\bf
E}_i)}^{\dagger}|e\rangle \langle e| (\widehat{\bf d}\cdot {\bf E}_j)|j,m'\rangle}
{\hbar^2\,[(\delta_L-kv-\omega_e)+i\gamma/2]} \langle j,m'|
+ \sum_{i\neq j,e,m,m'}|i,m\rangle \frac{\langle i,m|{(\widehat{\bf d}\cdot {\bf E}_j)}^{\dagger}|e\rangle
\langle e| (\widehat{\bf d}\cdot {\bf E}_j)|i,m'\rangle}
{\hbar^2\,[(\delta_L-kv+\omega_j-\omega_i-\omega_e)+i\gamma/2]} \langle i,m'|,
\end{equation}
\end{widetext}
where $\widehat{\bf d}$ is the optical transition dipole moment operator, $\delta_L$ is the laser (one-photon)
detuning, $\gamma$ is the optical linewidth, and $\hbar \omega_e$ is the energy of the excited-state HF levels.
Note that $\delta_L$ and $\omega_e$ are measured from a common zero level (hereafter from the HF level with
minimal momentum $F_e = I-J_e = 1$). In equations (\ref{Heff}) and (\ref{Rmatr}) the Doppler shift $kv$
($k=[k_1+k_2]/2$) and strong velocity-changing collisions are taken into account. The matrix $\widehat{R}$
contains resonant (first summation) as well as off-resonant (second summation) contributions to light shifts and
optical pumping rates (Hermitian and anti-Hermitian parts, respectively). The non-diagonal ($i\neq j$) elements
of the resonant term induce Raman coherence between the hyperfine levels of the ground state, which is
responsible for the CPT dark resonance.

Here we consider the steady-state regime, setting $(d/dt)\,\widehat{\sigma}_{gg} = 0$ in \eqref{maineq}. As a
spectroscopic signal, we take the total excited-state population $\pi_e$, which is proportional to the total
light absorption in optically thin media or to the total fluorescence. The following procedure is used to find
$\pi_e$. From \eqref{maineq}, the ground-state density matrix $\widehat{\sigma}_{gg}$ is expressed in terms of
$\pi_e$, and then $\pi_e$ is calculated from the normalization condition \eqref{n_cond}. In the general case of
arbitrary magnetic field this algebraic problem can be solved numerically.

At intermediate magnetic fields, when the magnetically sensitive
resonances are well separated from the magnetically insensitive
resonances, and simultaneously when the two magnetically
insensitive resonances ($-$1)--($+$1) and ($+$1)--($-$1) are well
overlapped (i.e. when their width is much greater than their
Zeeman splitting), the resonance lineshape is described with a
good accuracy by the generalized Lorentzian
\cite{DR_theory,KnappeAPB03}:
\begin{equation} \label{LLcpop}
\pi_{e} =C_0 - \frac{C_1\,(\widetilde{\gamma}/2)^2} {(\widetilde{\gamma}/2)^2+(\delta_R-\delta_0)^2} +
\frac{C_2\,(\delta_R-\delta_0)\,\widetilde{\gamma}/2} {(\widetilde{\gamma}/2)^2+(\delta_R-\delta_0)^2}.
\end{equation}
The coefficients in \eqref{LLcpop} ($C_i$, $\widetilde{\gamma}$, and $\delta_0$) can be analytically expressed
in terms of the parameters of the effective ground-state Hamiltonian (\ref{Heff}) in a way similar to the method
used in \cite{DR_theory}. Alternatively, they can be extracted from numerical results by a fitting procedure.

\end{document}